\documentclass[12pt,preprint]{aastex}

\def\sun{\hbox{$\odot$}}

\lefthead{}
\righthead{}

\begin{document}
\title{Absolute Proper Motions to $B\sim22.5$: IV.  Faint, Low Velocity White Dwarfs and
the White Dwarf Population Density Law}

\author{S. R. Majewski\altaffilmark{1} and M. H. Siegel\altaffilmark{2}}
\affil{University of Virginia, Department of Astronomy}
\authoraddr{P.O. Box 3818, Charlottesville, VA  22903-0818\\
e:mail: mhs4p@virginia.edu, srm4n@didjeridu.astro.virginia.edu}

\altaffiltext{1}{David and Lucile Packard Foundation Fellow; Cottrell Scholar of the Research 
Corporation}
\altaffiltext{2}{Present Address:  Space Telescope Science Institute, 3700 San martin Drive, Baltimore,
MD, 21218}

\begin{abstract}
The reduced proper motion diagram (RPMD) for a complete sample of 819 faint ($B \leq 22.5$) stars with 
high accuracy proper
motions ($\sigma_{\mu}\sim1$ mas yr$^{-1}$) in an area of 0.3 deg$^2$ in the North 
Galactic Pole field SA57 is investigated.  Eight 
stars with very large reduced proper motions are identified as faint white dwarf candidates.  On the 
basis of larger than 6$\sigma$ measured proper motions and the lack of photometric variability over a 
twenty year baseline, we discriminate these white dwarf candidates from the several times more numerous 
QSOs, which can potentially occupy a similar location in the RPMD.  For comparison, less than 
4$\sigma$ proper motions and photometric variability 
are found in all but one of 35 spectroscopically confirmed QSOs in the same field.  

While spectroscopic confirmation of their status as white dwarfs is a necessary, but difficult, 
outstanding task, we discuss the implausibility that these stars could be any kind of survey 
contaminant.  High quality proper motions
lend confidence in our ability to separate white dwarfs from subdwarfs in the RPMD.
If {\it bona fide} white dwarfs, 
the eight candidates found here represent a portion of the white dwarf population that hitherto has 
remained uninvestigated by previous surveys by virtue of the faint magnitudes and low proper 
motions of the stars.  This faint, low velocity sample represents an increase
in the white dwarf sky surface density to $B=22.5$ by an order of magnitude
over that found in the previously most complete surveys to this depth.  However, because the majority of 
the stars discovered
here are at projected distances of more than a disk scaleheight above the Galactic midplane,
their existence does not affect significantly the typical estimates of the local white dwarf density.
On the other hand, as distant white dwarf candidates with low, typically thin disk-like transverse 
velocities ($< 40$ km s$^{-1}$), the newly discovered stars suggest a disk white dwarf scaleheight 
larger than the values of 250-350 pc typically assumed 
in assessments of the local white dwarf density (and thought to characterize the Galactic old
thin disk in stellar population models).
Both a $<V/V_{max}>$ and a more complex maximum likelihood analysis of the spatial distribution
of our likely thin disk white dwarfs yield scaleheights of 400-600 pc while 
at the same time give a reasonable match to the {\it local} white dwarf volume density found 
in other surveys (although this good match is a result of the dominance of the one relatively nearby white dwarf in the
$1/V_{max}$ density calculation).  A high scaleheight persists even if the relatively small sample is 
pruned of any potential
thick disk or halo white dwarfs.

While our work is not optimized toward the study of halo white dwarfs as potential MACHO objects, our
results do have interesting implications for this hypothesis.  We can place some direct constraints (albeit
weak ones) on the contribution of halo white dwarfs to the dark matter of the Galaxy.  Moreover, the
elevated scale height that we measure for the thin disk could alter the interpretation of microlensing 
results to the extent of making white dwarfs untenable as the dominant MACHO contributor.
\end{abstract}
\keywords{}

\section{Introduction}

In Paper I of this series (Majewski 1992), proper motions were determined for nearly a 
thousand stellar objects in Selected Area 57 (SA57) at the North Galactic Pole to photographic 
$B_J \sim 22.5$ and $V_F \sim 21.5$.\footnote{$B_J$ is the passband produced by the combination of IIIa-J 
emulsion and GG385 filter, whereas $V_F$ is the combination of IIIaF + GG495.}
Photometric parallaxes were determined for a subsample of 250 stars with $0.3 \leq B-V \leq 1.1$ 
and $U \leq 21.5$
based on photographic ultraviolet excess measurements.  Since, in general, no direct measurement of 
the surface gravity of each star was readily available, a basic premise of the adopted analysis 
in Paper I
was that the survey stars are on the main sequence.  However, it is possible to exploit proper 
motions to discriminate luminosity classes of some stars through use of the reduced proper 
motion diagram (RPMD; see also Luyten 1922, Jones 1972a,b, Chiu 1980b, Evans 1992, Knox et al. 1999, 
hereafter K99; Cooke \& Reid 2000; Oppenheimer et al. 2001, hereafter O01).  
White dwarfs, on account of their very high reduced proper motions, should be readily identifiable in 
the RPMD.  This technique confers certain advantages over color searches for white dwarfs; 
e.g., it is possible to identify cool white dwarfs that are not distinguishable from the more
numerous late type field stars using colors alone.  Deep searches for faint, cool white dwarfs are 
important for testing white dwarf cooling models into the regime of Debye crystallization, and, by 
applying cooling theory in conjunction with the white dwarf luminosity function, to set limits 
on the star formation history and age of Galactic stellar populations. White dwarfs can also be 
used as tracers of the density laws of old populations, and white dwarfs are proposed as potentially
significant contributors to the dark matter component represented by 
gravitational microlensing events.

Numerous studies have attempted to establish the local density and/or luminosity function of 
Population I white dwarfs, and especially, recently, at the red end of the white dwarf sequence, 
due to the interest in cool white dwarfs for both age dating the Galaxy and as a primary 
source for microlensing candidates.  Results for
the derived local white dwarf density found among the different surveys still range by a factor of two 
(Fleming et al. 1986; Jahrei{\ss} 1987; Liebert, Dahn \& Monet 1988, 
hereafter LDM; Boyle 1989; Ruiz \& Takamiya 1995; Oswalt et al. 1996; 
Festin 1998; K99; Reid et al. 2001; Ruiz \& Bergeron 2001).  The question of completeness
lingers when considering the results of these various surveys.  So too
does a proper understanding of the density laws appropriate to the
samples garnered, since the conversion from a survey list to a local density 
requires an understanding of the {\it effective} volume surveyed,
i.e., the volume foreshortened by the drop-off in density with distance
from the Galactic midplane.  Typically, white dwarf studies have {\it adopted} a 
standard value for an exponential scaleheight of the disk in such
calculations, rather than attempted to {\it solve} for the density
law from their white dwarf samples.  This understandable reluctance derives from
the relatively limited range of distances probed by complete samples (typically 
one third to one half of the traditional old disk scaleheight),
which limits sensitivity to the form of the density law.  Survey incompleteness
at distances comparable to a disk scaleheight
derives both from photometric {\it and} astrometric limitations, since proper motions provide the 
most commonly used means by which to
identify white dwarfs (especially those redward of the field star main sequence turn off (MSTO)).  
Table 1 summarizes the major {\it astrometric}
white dwarf surveys to date (not including studies made from archival survey data, such as the Lowell
Proper Motion or Luyten Half-Second catalogues, which we represent by
the work of LDM).  With the 
exception of the deep, small area study by Chiu (1980b), these 
surveys are focused on stars with fairly large proper motions ($\gtrsim40$ mas yr$^{-1}$).  Such a 
limitation progressively excludes white dwarf populations with ever larger ranges of transverse 
velocity as a function of distance.  Figure 1 shows the limiting distances that are imposed 
on the detection of white dwarfs as a function of various apparent magnitude and proper motion
limits.  Figure 2 plots the sky density of detected white dwarfs against both photometric
and astrometric limits for the surveys listed in Table 1.

Table 1 and Figure 2 demonstrate that the detected white dwarf sky density appears to be more directly 
correlated with proper motion limits than survey depth.  This is an important point, one worth
considering given the new emphasis on properly accounting for the total white dwarf density in the
foreground of lensed sources.  The impressive K99 study, as the deepest, large area survey with the 
best proper motions to date, provides a benchmark for the present discussion.  K99
claim to find no evidence of incompleteness in their survey sample, and that this survey
represents the most complete large area sample to date is evidenced by their finding the largest sky 
surface density of white dwarfs for such a survey to date, 2.07 deg$^{-2}$.  Through a variety of 
arguments, K99 suggest that their proper motion limit of 50-60 mas yr$^{-1}$ provides a reasonable
compromise between minimizing spurious detections and maintaining completeness.   For example, 
for bright ($R \sim 14$) white dwarfs, K99 argue
that such a limit is more than enough to detect stars ``having a ({\it conservative})
tangential velocity of 40 km s$^{-1}$, [which] would have a proper motion of 
$\sim 80$ mas  yr$^{-1}$" (emphasis added); clearly for fainter, more distant ($> 100$ pc) 
examples of stars with similar transverse velocities, however their survey quickly becomes incomplete
(Figure 1).  It is worth noting at this point that the results of the present analysis will focus 
on the discovery of white dwarf candidates that are primarily {\it slower} than K99's ``conservative" estimate.  
An important means by which K99 attempt to build their case for completeness is by establishing that
their sample mean $(V/V_{max})$ statistic is nearly 0.5 (see \S 4), where here the $V$'s represent 
effective volumes {\it under the assumption of a 300 pc scaleheight}.  While K99's investigation
of variations in the assumed scaleheight show no significant alteration in their derived luminosity 
function, the authors do not state how varying the scaleheight affects their assumption 
of completeness.  $<V/V_{max}>$ is primarily a test of uniformity that K99 and others have adapted
to a test of completeness.  Even K99 admit that obtaining
$<V/V_{max}>=0.5$ cannot be regarded as proof of completeness, since an incomplete
sample can yield a similar result.  While this caveat might be considered all the more prescient
if it can be shown that 300 pc is not a proper scaleheight to assume in the effective volume
calculations in the first place, in the end, it should be noted that incompleteness in typical 
magnitude- and proper motion-limited samples may not significantly bias the derivation of the local 
luminosity function according to the Monte Carlo simulations of Wood \& Oswalt (1998) and 
Mendez \& Ruiz (2001).

In this article, we explore the question of the disk white dwarf population density distribution using
a low limit proper motion selected sample.  We employ reduced
proper motions to separate degenerate star candidates from subdwarfs and Population I main sequence 
stars within the deep proper motion sample of Paper I.  Of course, these proper motion techniques
have been used in many previous studies, some that explore to similar depth and/or much larger areas of 
sky than the Paper I sample (Table 1);  given our small survey area (0.3 deg$^2$), 
the volume we probe is much smaller than that explored by most previous surveys, even considering our
$B_J=22.5$ magnitude limit.  Within the last few years, several new surveys have also reached
$B\sim22.5$, and the K99 survey in particular has the potential, {\it based on photometric considerations},
to probe much larger volumes than our survey.  In practice, however, differences in {\it astrometric
quality} puts the present analysis in a unique niche in parameter space compared to all
previous studies:  No study in the literature has an astrometric precision comparable to
that afforded by the 16 year baseline study using 
deep photographic exposures on fine-grained emulsions with good plate scale discussed in Paper I. 
The resultant precision ($\sim1$ mas yr$^{-1}$ at $B_J=21.5$ and $\sim1.6$ mas yr$^{-1}$ at
$B_J=22.5$) allows for (1) an RPMD that
is relatively ``clean" of astrometric error-induced scatter at the white dwarf
locus, minimizing contamination problems, and (2) orders of magnitude smaller proper motion limits.  
Thus, we can find low proper motion white
dwarf candidates that would be missed by all previous surveys, look for white dwarfs at larger distances, 
and ensure a much higher level
of completeness than could be claimed before.  The transverse velocity limits
of a 1 mas yr$^{-1}$ survey to $B_J=22.5$ as a function of distance are shown in Figure 1;  
our astrometric advantage for probing to much larger distances than other surveys is clear.
As we shall show, this advantage allows us to find a white dwarf sky density at $B_J=22.5$
that is likely an order of magnitude larger than that found by K99, which itself had a density 
that was larger than found by most previous surveys (see Table 1).

The primary contribution to our sky density seems to be
from an extended distribution of distant white dwarfs with cold kinematics typical of the
thin disk.  That the majority of the white dwarfs have projected distances larger
than 300 pc, and a third of them beyond 900 pc, suggests the need for a revision of the normally
assumed 250-350 pc scaleheight for the thin disk white dwarf population.
Thus, a principal endeavor in this contribution is to use this admittedly small, but much less 
kinematically biased, sample of white dwarfs
to define better the vertical distribution of the old disk, a task to which our unusually distant
sample has particular leverage.

\section{Identification of White Dwarf Candidates}

We define the $F$ passband reduced proper motion of a star, $H_F$, as 

\begin{equation}
H_F = F + 5\textrm{log}\mu + 5 = M_F + 5\textrm{log}V_T - 3.378
\end{equation}

\noindent where $F$ is the calibrated photographic flux on IIIa-F emulsions with a GG495 filter, 
$M_F$ is the absolute magnitude of a star, $\mu$ is the proper motion
in arc-seconds per year and $V_T$ is the transverse velocity in km s$^{-1}$.  Because 
the properties of the stars in a given population should cluster around certain values, stars can 
be classified based on their location in the RPMD.  However, as demonstrated by Chiu (1980b), 
many stars may not be assignable to a unique population/luminosity class on the 
basis of the RPMD.  The ridge lines of different populations intersect because the 
combined properties of populations according to equation (1) overlap.  Fortunately, we explore 
here a comparatively simple region of the RPMD, the extreme domain typically inhabited by white 
dwarfs.	

The sample of SA57 objects for which proper motions and new photographic photometry were calculated 
in Paper I were those having $B_J \leq 22.5$ or $F \leq 21.5$ in the catalogue of Kron (1980) (where
the stated magnitude limits are based on Kron's photometry from the same first epoch
plates and $B_J=J$ refers to the IIIa-J emulsion with GG 385 filter, and $B_J\sim B$).  Figure 3 shows 
the ($J-F, H_F$) RPMD for this sample, excluding confirmed QSOs and galaxies (see M91; Travese 
et al. 1994, hereafter T94).  
The photometry in this diagram is taken from an analysis using more photographic plates as well 
as a more extended list of photometric standards than is described in Paper I.  Typical random errors 
in $J-F$ for these data are $\sim0.02$ magnitudes but reach 0.10 magnitudes at the magnitude limit 
(see Figure 3 of Paper I).  Random errors in $H_F$ are typically 0.2 magnitudes but in some cases 
exceed 1.5 at the magnitude ({\it not} $H_F$) limit (see Figures 2 and 10 in Paper I).  These stars 
with large error are almost entirely stars with low $H_F$ and do not affect our analysis of high 
$H_F$ objects (see Figure 3) since none are near the WD1 locus.

Except for that marked ``WD1", the ridge lines shown in Figure 3 are those from Chiu (1980b), 
converted to the photographic ($J-F, H_F$) system using the transformation equations given in Table 5 
of Paper I.  The six populations shown are the Population I main sequence (MS1), red giants (RG1) and 
white dwarfs (WD1) and the Population II subdwarfs (SD2), red giants (RG2) and horizontal branch (HB2).  
Chiu made no provision for an Intermediate Population II (i.e., a ``thick disk"), but the ridge lines 
for such a population would presumably lie somewhere between the Population I and Population II ridge 
lines.  We refer the reader to Chiu (1980b) for details of the derivation of the ridge lines, but 
review and revise here his formulation of the WD1 ridge line since it is particularly germane to the 
present discussion.  Chiu's WD1 color-magnitude sequence combines the calibration for DA dwarfs by 
Sion \& Liebert (1977) for stars blueward of $B-V = 0.4$ and by Greenstein (1976) for stars redward 
of $B-V = 0.4$.  In order to compute the RPMD ridge line for the WD1, the kinematics adopted were for a 
standard solar motion of 19.5 km s$^{-1}$ and a velocity ellipsoid ($\sigma_u, \sigma_v, \sigma_w$) = 
(35, 25, 20) km s$^{-1}$.  This yields a peak in the $V_T$ distribution at 44 km s$^{-1}$.

We have updated Chiu's WD1 locus using an empirical
color-absolute magnitude relation derived from the
recent catalogue of white dwarf parallaxes published by Bergeron, Legget \& Ruiz (2001).  
Corrected for Lutz-Kelker (1975) bias and cleaned of white dwarfs with poor photometry
or $\sigma_{\pi}/\pi > 0.2$, this produces the relations:

\begin{eqnarray}
M_V = 11.76 + 5.191 (B-V) - 1.615 (B-V)^2 &  \nonumber\\
M_F = 11.65 + 5.132 (J-F) - 1.771 (J-F)^2   &   (-0.4 \leq J-F \leq -0.05)\\
M_F = 11.71 + 4.057 (J-F) - 1.164 (J-F)^2 &     (-0.05 \leq J-F \leq 1.5) \nonumber
\end{eqnarray}

\noindent This color-magnitude relationship is very similar to the Greenstein relation for cool
white dwarfs and while Bergeron et al. do not have any hot white dwarfs, the blue extrapolation
of this curve is similar to the Sion \& Liebert derivation.  This relation is an empirical
one derived from the entire Bergeron et al. sample and does not distinguish between DA and non-DA
stars.  However, subsampling by white dwarf type does not significantly alter the color-magnitude 
relationship, nor does it reduce the dispersion about the best fit relationship.  Given the lack of 
spectra with which to classify the white dwarf candidates we identify here as either DA or non-DA, we believe that the 
safest method of analysis is to use this empirical relationship for both photometric parallaxes 
and the derivation of the RPMD ridge line.  The dispersion in the fit ($\sigma_{M_V}$=0.5) to the color-magnitude
relationship is about twice 
as large as the dispersion in the color-magnitude relationships used for main sequence stars by Siegel 
et al. (2001), resulting in less reliable photometric parallaxes in the white dwarf case.  
However, the dispersion is smaller than the 0.8 magnitude dispersion in the WD1 color-magnitude relationship 
used by Chiu and therefore represents an improvement over the latter.

The corresponding WD1 ridges lines, adopting Chiu's WD1 kinematics, are:

\begin{eqnarray}
H_V = 16.60 + 5.191 (B-V) - 1.615 (B-V)^2 & \nonumber\\
H_F =  16.49 + 5.132 (J-F) - 1.771 (J-F)^2   &   -0.4 \leq J-F \leq -0.05\\
H_F =  16.54 + 4.057 (J-F) - 1.164 (J-F)^2   &   -0.05 \leq J-F \leq 1.5 \nonumber
\end{eqnarray}

From Figure 3 it can be seen that the majority of very red stars lie along the Population I main 
sequence whereas the majority of the bluer stars tend to clump along the SD2 and lower RG2 ridge lines.  
This is to be 
expected since redder stars are mainly nearby, late type dwarfs from the old disk while the bluer stars 
tend to be several kpc above the Galactic plane and are therefore members of the Intermediate 
Population II or halo.  Figure 3 also shows a smattering of stars along the upper red giant loci as well 
as along the WD1 sequence.  The main population overlap in the lower part of the RPMD is due to 
scatter between the WD1 and SD2 populations.

Figure 4 highlights the SD2/WD1 region of the RPMD.
On the basis of a much larger sample of stars, Jones (1972b) suggested that a good dividing line 
between the WD1 and SD2 populations is $H_V = 6.84 (B-V) + 13.0$, which can be translated to the 
photographic system with the relations in Table 5 of Paper I as $H_F = 5.83 (J-F) + 12.93$.  Using this division 
(dotted line in Figure 4) leaves eleven white dwarf candidates: Seven near the Chiu WD1 locus and 
four strung along and just below the Jones division line near $J-F \sim 0.3$.  We have also added
object (17424), located at $(J-F, H_F) = (1.25, 18.5)$, the nature of which we will discuss in \S3.

We assemble all available observational data for these twelve objects
in Tables 2 and 3.  Table 2 gives photometric data: Column 1 
gives the Kron (1980) catalogue identification, columns 2 and 3 give the $F$ magnitude and its 
error, columns 4 and 5 give the $J-F$ color and its error, columns 6 and 7 give the $U-J$ color 
and its error, and column 8 gives the variability index, v.i., as defined in M91, and where 
variability may be taken as v.i. $>$ 1.75.

Table 3 gives the astrometric data for the objects: Columns 2 and 3 give the right 
ascension and declination of each object for equinox J2000.0 in epoch 1989.97, columns 4 and 5 give the proper 
motion in the direction of the Galactic anticenter ($l = 0^{\circ}$) and its error, column 6 and 
7 give the proper motion in the direction of Galactic rotation ($l = 90^{\circ}$)\footnote{Since we
are working very near to the North Galactic Pole, where the traditional $\mu_l$ and $\mu_b$ are not defined, 
we adopt the NGP descriptions $\mu_{l=0}$ and $\mu_{l=90}$.}, columns 8 and 
9 give the reduced proper motion, $H_F$, and error, and column 10 gives the proper motion index 
(defined as the total proper motion divided by the total proper motion error).

While the seven stars near the WD1 locus would seem securely identified as white dwarfs on the 
basis of both the updated Chiu ridge line and the Jones criterion, as well as their predominantly 
blue colors (at least five and perhaps six are blueward of the SD2 MSTO), the classification of 
the latter four objects is more ambiguous.
Because the Jones division is based on a much brighter sample of stars, its appropriateness might 
be questioned in the context of our very deep sample, which likely contains subdwarfs 
with more extreme kinematics than represented either by Chiu's SD2 ridge line or Jones' subdwarf 
sample (which is derived from the Yale Bright Star Catalogue).  For example, among our subdwarfs are 
members of the ``moving group" discussed in Majewski, Munn \& Hawley (1994, 1996), the RPMD of 
which is shown in Majewski (1999), where a $V_T=245$ km s$^{-1}$ ridge line is found to give a good
fit.  It is therefore worthwhile to demarcate from general principles 
the limit of the domain accessible to the SD2 population.  First, we note that the Chiu 
color-magnitude relation for the SD2 population is identical to that of the MS1 population, but 
reduced in magnitude by 0.85 magnitudes.  Based on the results of the U.S. Naval Observatory 
parallax program (Monet et al. 1992), it is known that metal poor stars may be subluminous by as 
much as 2.5 magnitudes compared to their solar metallicity counterparts.  The dot-dashed line in 
Figure 4 is the SD2 sequence increased first by 1.65 magnitudes, and then by an additional 
1.75 magnitudes to account for the increase in 5 log $V_T$ as a result of increasing $V_T$ to the 
maximum transverse velocity accessible to any star (from the modal $V_T$ value of 315 km s$^{-1}$ 
used by Chiu).  A reasonable limit to the maximum velocity of an SD2 star is taken to be the 
Galactic escape velocity in the solar neighborhood, 475 km s$^{-1}$ (Cudworth 1990), with 
the addition of the up to 231 km s$^{-1}$ motion of the Sun in the Galactic rest 
frame\footnote{Note that this limit in $V_T$ is not taken into account in Chiu's formulation of 
the 5 log $V_T$ distribution as shown in his Figure 2b.} that can be added as a reflex motion, so 
that $V_T$=475+231=706 km s$^{-1}$ 
represents the maximum heliocentric velocity for a star bound to the Milky 
Way in the solar neighborhood.  The dot-dashed line in Figure 4 marks 
this absolute limit of main sequence stars in the RPMD. In addition, the MSTO for 
Population II, which is somewhere near $J-F \sim 0.4$, places an additional constraint on the 
domain of the SD2 population.  From these arguments, we conclude that the four points along and 
just below the Jones (1972b) division in Figure 4 are not members of the Galactic SD2 population.

It is also possible to establish reasonable limits on the vertical extent of the white dwarf 
population in the RPMD.  No hard {\it minimum} limit to $H_F$ may be established for any stellar 
population since any star may have arbitrarily small $V_T$.  However, based on the 5 log $V_T$ 
distribution for Population I stars as calculated by Chiu (see his Figure 2a), a 90\% limit for 5 
log $V_T$ is $\sim5.2$ and a 75\% limit is $\sim6.2$.  These kinematical limits are indicated by the 
dashed lines in Figure 4.  It can be seen that the sequence of four points at $(J-F, H_F) \sim 
(0.3, 16)$ fall between the 75\% and 90\% limits of the WD1 sequence.  As there are only seven
highly likely white dwarfs (those nearer the WD1 ridge line), 
it is improbable that more than a couple of white dwarfs would occupy the extreme low velocity
($V_T \leq 14$ km s$^{-1}$) wing of the velocity dispersion and the extreme low $H_F$ white
dwarf limit.
Thus, it may be concluded that most of the four objects along and just below the Jones 
(1972b) division line are neither white dwarfs nor subdwarf stars.  Based on the 
475+231 km s$^{-1}$ limit on the tangential velocity, the possibility that these objects could be 
either in the HB2 or the MS1 populations is also excluded.  We will now argue that these four
objects are, in fact, most likely to be QSO's.  

\section{QSO Decontamination}

An additional application to which the deep photographic plates discussed in Paper I have been applied 
is the search for faint QSOs (Koo, Kron \& Cudworth 1986, hereafter KKC; Koo \& Kron 1988, hereafter KK88; 
T94) 
using a variety of techniques.  The high precision astrometry from Paper I, as well as the multi-epoch 
photometric data produced therein, allow a new search for QSOs using the criteria of lack of proper 
motions and photometric variability (Majewski et al. 1991, hereafter M91; T94).  Proper motion data are 
particularly effective at identifying a major contaminant in color-based QSO searches -- white 
dwarfs -- and it was the decontamination of the QSO sample that partially motivated the present 
study (M91).  However, the contamination of deep, low proper motion white dwarf 
samples, such as that discussed here, {\it by QSOs} is a potentially more insidious problem
given that QSOs outnumber white dwarfs by 5:1 to $B=22.5$.

We include in Figure 4b the location of spectroscopically confirmed QSO's in the survey conducted 
in this field by Kron et al. (1991) and T94.  As can be seen, several spectroscopically confirmed 
QSOs inhabit the same location in 
the RPMD as the four questionable objects.  In addition, unlike the seven more likely white 
dwarfs that fall along the WD1 locus, all of which have very definitely measured proper motions 
(i.e. greater than $6 \sigma$), the four objects in question all have measured proper motions 
less than three times their error (Table 3).  

A more complete analysis of the combined proper motion and variability characteristics of stellar 
objects in this SA57 sample is discussed in M91 and revisited in T94.  In both papers, which have
independent photometric analyses, all but one of the 35 known QSOs in SA57 are found to be variable within 
the same multi-epoch plate collection (M91) or a subset thereof (T94).  M91 discuss 
the effectiveness of separating QSOs from other stellar objects in a variability-proper motion
diagram (see, e.g., Figure 2 of M91).  Using the definition of ``variability index (v.i.)" and 
``proper motion index (p.m.i.)" (essentially total proper motion divided by its uncertainty) defined in M91, 
where it is found that QSO's almost entirely have both p.m.i. $<$ 4.0 and v.i. $>$ 1.75, we find that
three out of four of the questionable white dwarf candidates satisfy the same criteria.  The fourth 
object (9361) is not variable, but does have p.m.i. $<$ 3.0.  These properties are characteristic of a 
compact narrow emission line galaxy (see M91).  Finally, the positions of these four 
questionable objects in the RPMD is so similar to the location of known QSOs, we consider it
highly probable that all four are QSOs (or compact galaxies) and consider them as such for the remainder of 
the paper.

The above discussion should raise a note of caution for those using the RPMD to find white dwarfs at 
faint magnitudes.  QSO's appear to be the an important contaminant of the white dwarf region in the RPMD, 
especially when the Jones criterion is used.  This problem can be avoided
by only selecting targets with a certain minimum p.m.i. ($\mu/\epsilon_{\mu}$) -- assuming that 
no uncorrected, systematic astrometric effects cause QSO's to have a spurious motion (for example, 
color errors are a common problem in QSO proper motion measurements because QSOs have very different 
spectral energy distributions from stars, see, e.g., Paper I).  Such a proper motion selection, of course, 
induces a kinematical bias in any analysis of the density distribution or dynamics of white 
dwarfs.  As can be seen from Figure 1, however, a p.m.i. selection bias is much less severe for our
$\sim1$ mas yr$^{-1}$ limited survey compared to the 50 mas yr$^{-1}$ surveys.  Even at 1 kpc distance, 
our p.m.i. = 6 cutoff only excludes stars with $V_T \leq 28$ km s$^{-1}$.

We call attention to one final object in Figures 3 and 4, located at $(J-F, H_F) = (1.25, 18.5)$.  This 
object (17424) falls midway between the WD1 and SD2 ridge lines but well below the 75\% WD1 line.  
Jones would not have selected this object as a white dwarf but, again, his sample was biased against 
finding stars in this part of the RPMD and his selection criterion is a reflection of that
fact.  Spectroscopy of this object would be useful since the 
number of intrinsically faint white dwarfs, which determine the faint end of the luminosity 
function, is small and it is these few white dwarfs that are critical to constraining the age of 
the Galaxy (see, e.g. Wood 1992; Oswalt et al. 1996; K99 and references therein) through 
white dwarf cooling theory.  Such cool white dwarfs may also be the paradigm MACHO (MAssive Compact
HalO) objects.
Redder white dwarfs have been found as members of common proper motion pairs in proper motion selected 
surveys (see, e.g., Hintzen et al 1989; Silvestri et al. 2001) and, recently, in large area surveys 
such as O01 and K99.  If 17424 is a true white dwarf, it will have been found without 
the usual degree of selection bias and it lies nearly twice as far away as the most distant cool white 
dwarfs in O01.  Even if it is not a white dwarf, object 17424 would be a curious star -- either 
very subluminous or very high velocity, or both.  We note that it is definitely a star, as it 
has a 13.5 $\sigma$ proper motion measurement.

This leaves us with seven high probability white dwarf candidates, plus one possible red 
candidate which, however, would have been excluded by the Jones (1972) criterion.  In comparison, 
at least 35 QSOs have been confirmed in this field (KK88; M91; Majewski et al. 1993; 
T94) and more are expected to be confirmed (even excluding the four possible QSOs discussed
above -- see M91), so that QSOs outnumber white dwarfs by at least a 
factor of five to the magnitude limit of this survey.  In contrast, the ratio of QSO's 
to white dwarfs in the much brighter Palomar Green survey is 1:5.

In the course of their QSO search using a subset of the plate material used here, KKC,
KK88 and T94 identified white dwarfs in the SA57 field based on the combination 
of color, proper motion, and variability criteria.  The present analysis represents an improvement 
in the candidate selection in KKC by (1) significantly more reliable astrometry (see the comparison 
of the Paper I astrometry to the KKC astrometry in Paper I), (2) an improved variability analysis using more plates covering 
more epochs, and (3) an improved photometric analysis (again, even over T94) using more standard 
calibrators as well as more plate measures.  KKC identify eight white dwarf candidates in 
SA57, among them the stars 5003, 10347, 13612 
and 13786 also identified as white dwarfs here (KKC numbers 71, 65, 76 and 69 respectively).  Of the 
remaining white dwarf candidates KKC identify, subsequent spectroscopy (KK88) has revealed object 
10028 (number 1 in KKC) to be a QSO, object 19387 (number 51 in KKC) to be a likely QSO, and object 
11334 (number 12 in KKC) to be a possible main sequence star.  The remaining white dwarf candidate 
in KKC, object 710 (number 62 in KKC) is (according to the analysis of M91) a likely QSO on 
the basis of no measured proper motion as well as variability.  KKC identify our present white dwarf 
candidate 10405 (KKC number 24) as a possible narrow emission line galaxy, but this object clearly 
has a proper motion (Table 3).

\section{White Dwarf Density Law}

Although the number of candidates in the SA57 survey field is small, it represents one of the 
deepest existing samples of white dwarfs -- likely to be complete to $B_J$ = 
22.5, a flux limit comparable to the K99 study.  Five of the eight candidates are fainter than the 
$F$ = 19.5 limit of the search by O01, and four by more than a magnitude.  Far more importantly, 
though, our data represent the most {\it astrometrically} complete survey in
the literature, with a proper motion precision six times better than the previous best (K99) at
this magnitude limit and to a depth 2-2.5 magnitudes fainter than the the only comparable
astrometric study (Chiu 1980a).  At 
the risk of overinterpretation of a meager sample, it is of interest to consider these data as a 
probe of the white dwarf population density law since these stars provide unprecedented leverage at 
relatively large  distances (greater than a scale height) from the Galactic plane.  

Table 4 presents derived properties of the white dwarf sample.  The $J-F$ magnitudes were converted 
to $B-V$ using the conversions of Paper I.  The absolute magnitude $M_V$ was then determined from
our revised $M_V(B-V)$ relation (\S2) and then converted back to $M_F$.  
Columns 10 and 11 of Table 4 give the derived $M_F$ and $M_V$ for each 
white dwarf candidate.  Column 12 gives the cooling time for pure hydrogen models, 
as calculated from the absolute magnitudes using tables 1 of Bergeron et al. (1995).  These
ages are consistent with thin or thick disk membership for our white dwarf candidates and are clearly
not old enough to be part of some primordial stellar population.
 
From the derived absolute magnitude, the photometric parallax and its error are formulated 
(columns 2 and 3 of Table 4).  With the derived distances, the proper motions may be converted to $U$ 
and $V$ velocities, which have been formulated using the basic solar motion, $(u_{\sun}, v_{\sun}) = (-9, 11)$ km 
s$^{-1}$.  Note that because SA57 is not precisely at the Galactic pole, but at $b=86^{\circ}$, 
there is a small (0.7\%) velocity uncertainty due to the lack of measured radial velocities -- 
which leaves an uncertain, but likely minute component of radial velocity that contributes to $U$ 
and $V$.  The derived values of $u=U+u_{\sun}$ and $v=V+V_{\sun}$ and their errors are given in 
columns 4-7 in Table 4.  For the majority of the 
stars, the kinematics appear to be those of the old disk, with $(u^2 + v^2)^{1/2} < 40$ 
km s$^{-1}$ (column 9).  However, stars 10347 and especially star 5003 have more extreme 
kinematics, with $u$ velocities more like that expected for the Intermediate Population II or thick disk. 

The interesting aspect of the derived velocities for our white dwarf sample is that they show that we are
finding a significant {\it low velocity} population of presumably ``thin" disk WD's at large 
distances.  {\it That the majority of these stars have $V_T < 40$ km s$^{-1}$ shows this is a subset
of white dwarfs hitherto uninvestigated by K99 or any other survey}.  The eight stars have a 
remarkably cold kinematical signature of ($<u>,<v>$)=(25,-12) and ($\sigma_u,\sigma_v$)=(46,22) km sec$^{-1}$.  
Removing the high velocity star 5003 changes this distribution to ($<u>,<v>$)=(13,-10) and
($\sigma_u,\sigma_v$)=(32,22) km sec$^{-1}$.  This compares well to the 
($\sigma_u,\sigma_v$)=(35,25) km sec$^{-1}$ velocity dispersion
of late-type Population I stars taken from the McCormick spectroscopic survey (Chiu 1980a) or the
more recent ($\sigma_u,\sigma_v$)=(35,21) km sec$^{-1}$ velocity dispersion of 
``component A" nearby M dwarfs derived in the survey of Reid et al. (1995).
Thus, we see that while kinematical bias
is not completely eliminated by our proper motion selection, it is severely reduced (e.g., only losing
stars with $V_T < 14$ km s$^{-1}$ at 500 pc).  
If the kinematics of the sample so closely reflect the previously derived kinematics of the thin
disk, there is reason to believe that the spatial distribution we derive is similarly unbiased.

The ``interim" starcount model described by Reid \& Majewski (1993) predicts that 4.4 white dwarfs 
should be present in the 0.3 deg$^2$ field of SA57 at the imposed magnitude limit.  More than 90\% 
of this predicted count is from the old thin disk.  It is interesting, if only marginally 
significant, that we find almost double this predicted number of white dwarf candidates.  However, 
it is possible that the higher discovery density is a result of a higher white dwarf scale height 
than the 325 pc height utilized for the old thin disk of the ``interim model".  Subsequent work on
refining these models (Siegel et al. 2001) has derived a lower thin disk scale height generally, but shows
that fainter, late-type dwarfs have higher thin disk scale heights.

Analysis of the density law from photometric parallax must take into account the effects of
Malmquist bias (Malmquist 1920).  Most studies that attempt to measure completeness or local
density do not correct for this bias and this could systematically affect their analyses. 
Malmquist bias is very strong for white dwarfs because of the
high dispersion of the color-magnitude relation.  Using Malmquist's 
formulation, this results in a standard correction to the absolute magnitudes of 0.18 
given a dln(A)/dm slope of 0.7.\footnote{In our study of the photometric parallax of
main sequence stars (Siegel et al. 2001), we have found that dln(A)/dm is a function of the absolute
magnitude of the sources.  The slope is very shallow for blue main sequence stars (dln(A)/dm$\sim0.35$).  
For our red Population I main sequence 
stars, however, we find a value of $\sim$0.7, which is slightly lower than the Chiu (1980b) 
estimate of dln(A)/dm $\sim$ 0.8 for WD1.  A more conservative (smaller) Malmquist correction
lowers the derived scaleheights only modestly.}  It bears repeating
that the Malmquist correction applies only when a sample of stars is analyzed in a statistical
sense.  It is inappropriate to apply this correction to any particular star on its own.  Thus the
individual distances in Table 4 are not corrected for Malmquist bias.  The following analysis of the
density law, however, utilizes Malmquist-corrected distances.

In order to explore the white dwarf density law, we have utilized two statistical tools.  The first is the $V'/V'_{max}$ 
technique of Schmidt (1968, 1975).  A good description of the technique may be found in Chiu (1980a).  The definitions of 
the ``weighted volume elements" $V'$ and 
$V'_{max}$ are given by

\begin{equation}
V' = \omega \int_0^{r_{WD}} D(r)r^2dr
\end{equation}

\noindent and,

\begin{equation}
V'_{max} = \omega \int_0^{r_{max}} D(r)r^2dr
\end{equation}

\noindent where $r_{WD}$ is the photometric parallax distance to a white dwarf and where $r_{max}$ is 
the maximum parallax distance 
to which the white dwarf could have been observed.  It is an important but sometimes ignored point that
this limit can be photometric {\it or astrometric}, and the latter is often the more stringent.  
In the case where $r_{max}$ is imposed by the photometry limit, we assign $r_{max}$ for each star as 
the distance that it would have with the same absolute magnitude, but an apparent magnitude at the 
limit of the survey, given by $F=21.5$ or $J=22.5$, whichever yields the most distant photometric parallax.  For the 
astrometric limits, 
rather than assuming a model velocity distribution, we 
take $r_{max}$ for each star as the distance at which the proper motion would fall below 
3$\sigma$ given its transverse velocity.  $r_{max}$ is therefore defined as:

\begin{equation}
r_{max} = min[r_{WD} \frac{p.m.i.}{3} , max[r 10^{0.2(21.5-F)},r 10^{0.2(22.5-J)}]]
\end{equation}

\noindent Column 8 in Table 2 gives $r_{max}$ 
for each white dwarf.  The assumed disk density law is $D(z) = \phi \exp((z_{\sun}-|z|)/z_0)
\exp((r_{\sun}-r)/r_0)$ where 
$z_0$ is the scaleheight, $r_0$ is the scale length, $z_{\sun}$ and $r_{\sun}$ are the solar
height and radius, respectively and $\phi$ is the local density normalization.  The parameters
of $r_0$, $r_{\sun}$ and $z_{\sun}$ are taken from Siegel et al. (2001).

The values of $<V'/V'_{max}>$ as a function of scaleheight have been calculated for a variety of 
subsamples, as summarized in Table 5 and Figure 5.  The mean value of $V'/V'_{max}$ should be 0.5 
when the correct density law is adopted.  It should be noted that the cool white dwarf candidate 
17424, which may actually be an extremely metal-poor, high velocity subdwarf, is near the magnitude 
limit of the survey.  It therefore has a high $V'/V'_{max}$ and has a large influence on the 
calculation of $<V'/V'_{max}>$ for our small sample.  Therefore, families of curves are shown in 
Figure 5 both including this object as a white dwarf (solid lines) and excluding it (dotted lines).  

An important assumption implicit in adopting the above density law is that a {\it single} such density law 
applies.  However, when probing to deep magnitudes, it is possible that a non-negligible fraction 
of the white dwarf sample may be from Population II.  As discussed above, two of the white dwarf 
candidates in Table 2 have somewhat more extreme kinematics, which may identify 
them as thick disk or halo members.   In addition, star 17424, which
dominates the $1/V'_{max}$ density calculation, is redder, cooler and thus possibly older than the 
bulk of our stars and is the most likely, based on color, to be part of an old population (although
it is kinematically cold in $u$ and $v$ velocities).
Thus, we have tested subsamples with various of the white dwarfs (Table 5) 
excluded under the assumption that they are not members of the thin disk (either on the basis of 
distance or kinematics) or not white dwarfs at all (i.e. 17424) and we have sought the best 
fit $<V'/V'_{max}>$. 

The largest error in the calculation of $<V'/V'_{max}>$ is the sampling error, given by 
$1/(12N)^{1/2}$ (Chiu 1980a), where $N$ is the number of stars in the calculation, and is 
0.12-0.10 for $N$ = 6-8.  Thus, a reasonable range of $<V'/V'_{max}>$ that 
might be expected when the correct density law has been adopted is 0.38-0.62.  From Figure 5 and 
Table 5 it can be seen that with object 17424 excluded, the various subsamples give 
$<V'/V'_{max}>$ = 0.5 at $z_0 >$ 550 pc, but a large range of scaleheights -- 300 pc to greater than 
1000 pc -- is acceptable.  When star 17424 is included, $<V'/V'_{max}>$ = 0.5 occurs at $z_0 >$ 
700 pc, and the minimum ``reasonable" scaleheight (when $<V'/V'_{max}>$ = 0.62) is about 350 pc.  
These exceptionally large scaleheights are in conflict with the more modest white dwarf 
scaleheights typically derived -- for example, Boyle's (1989) scaleheight of $275 \pm 50$ pc -- 
although Chiu (1980a) 
obtained similarly high (400-500 pc) scaleheights for his relatively deep ($V < \sim 20.5$) 
sample of white dwarfs.  Recent results from starcount analyses of non-degenerate stars
include thin disk scale heights of 
280 pc (Siegel et al. 2001), 330 pc (Chen et al. 2001), 290 pc (Buser et al. 1999), 240 pc (Ojha et al. 1999) and 325 pc 
(Larsen 1996).  Ng et al. (1997) parameterize the thin disk as having three components
of young, intermediate and old age;  the old thin disk population in their model has a scale height
of 500 pc but is not well-constrained in their analysis.

The method of maximum likelihood provides a different likelihood estimation than the 
$<V'/V'_{max}>$ method.  The method of maximum likelihood is generally favored in any analysis of 
a small sample of data points because of it's unique sensitivity (Bevington \& Robinson 1992).  The 
likelihood function in this particular application would defined by:

\begin{equation}
\mathcal{L}\mathit = \prod_i \frac{D(r_i)r_i^2}{\int_0^{r_{i,max}} D(r) r^2 dr}
\end{equation}

\noindent Although this function gives the appearance of having units of pc$^{-1}$, the numerator
is intrinsically integrated over a delta function for the distance of each white dwarf
(uncertainty having already been corrected from Malmquist bias).  This unseen delta function
produces a dimensionless likelihood.
The maximum likelihood is then normalized to the posterior probability by integrating
the probability distribution over a uniform prior probability distribution with scaleheights of 
$200 < z < 1000$ (a reasonable range of possible values).  The most likely scaleheight values 
obtained are 
shown in the last column of Table 5 and illustrated in Figure 6.  In this method, stars are not 
assigned to a particular population but are given a probability distribution based on their spatial 
position and the distribution of the relevant populations.  Thus, a full model including a thick disk and 
halo populations, as well as radial density variations as defined in Siegel et al. (2001) is 
appropriate.  Only the thin disk scale height was varied since the likelihood proved very insensitive 
to thick disk parameters and completely insensitive to halo parameters.

The method of maximum likelihood should be less sensitive to small number statistics
than is $<V'/V'_{max}>$.  As may be seen, the parameters derived from
various subsamples of the white dwarf candidates yield smaller scaleheights, a smaller overall 
range of results, and the star 17424 has a 
far smaller impact upon the derivation of the scaleheight (although it significantly alters the local
normalization).  Nevertheless, as may be seen, large scaleheights ($z_0 > 380$ pc) are still found
with this method and the range of acceptable scaleheights is still quite large.  If we take
the 2$\sigma$ uncertainty in the scaleheight to be the point at which the likelihood is half
its peak value, possible maximum likelihood scaleheights range from 220 to 890 pc.

If our candidates are actual white dwarfs (and the data for all
but 17424 strongly suggest that they are), they increase the sky density
of white dwarfs to $B=22.5$ to more than 20 deg$^{-2}$ (as high as 27 deg $^{-2}$ if all eight 
candidates are real white dwarfs), an order of magnitude larger than that found by K99, the previously
most complete white dwarf survey to date.  Given the increased sky density provided by this population,
one might expect a significant impact on the derived local white dwarf density and luminosity 
function.  However, as Wood \& Oswalt (1998) and Mendez \& Ruiz (2001) have argued, incompleteness in white dwarf surveys
does not impact estimates of the local luminosity function or density.  We have shown that their 
analysis, based on Monte Carlo simulations,
is correct, at least as it pertains to the present study.  Our sample represents almost 
exclusively a {\it distant} ($z > z_0$) population.  Table 5 summarizes the local white dwarf
density associated with each curve in Figure 5 on the basis of the $\sum_i (1/V'_{max,i})$ 
technique for each scaleheight derivation.  As can be seen, extrapolation
of our density laws to $z=0$ provides local densities in the range of values obtained by other 
groups (see below).  Here, however, our results succumb to the vagaries of small number 
statistics.

Fleming, Liebert \& Green (1986) derived a local white dwarf density of 0.49 $\pm$ 0.05 
per 1000 pc$^3$ for white dwarfs with $M_V <$ 12.75, while Boyle (1989) derived 0.60 $\pm$ 0.09 
per 1000 pc$^3$ for $M_V <$ 12.75.
We provide data for our five stars with $M_V <$ 12.75 in Table 5 and Figure 5 (dashed curve).  For 
these intrinsically brighter white dwarfs we do obtain a lower scaleheight, 450 pc, but this value
is still higher than Boyle's.  Our derived local density for $M_V <$ 12.75 white dwarfs is consistent 
with both previously named studies, but only when the higher scaleheight is used; adopting Boyle's 
lower scaleheight results in a significantly higher local density, inconsistent with other surveys.  
For the entire sample, our local density is very close to the $3.95 \times 10^{-3}$ pc$^{-3}$ derived
by Reid et al. (2001) from a sample of local stars, the estimated density from
K99 of $4.16 \times 10^{-3}$, and the density of Ruiz \& Bergeron (2001) of $5.6 \times 10^{-3}$.

\section{Relevance to MACHO's}

Early micro-lensing analyses suggested that half of the dark halo of the 
Milky Way could be comprised of 0.5 $M_{\sun}$ MACHO objects (see, e.g., Alcock et al. 1996).  This 
fractional contribution has recently been revised downward to approximately 20\% of the dark 
matter (Alcock et al. 2000; Lasserre et al. 2000).  
Because deep starcounts surveys have shown that low-mass red dwarfs
cannot be a major contributor to the dark mass (Flynn, Gould \& Bahcall 1996; 
Reid et al. 1996; Santiago, Gilmore \& Elson 1996; Gould, Bahcall \& Flynn 1997), white 
dwarfs have been
cited as a possible source (cf. Charlot \& Silk 1995) -- indeed, ``the least unlikely candidates" 
(Mera, Chabrier \& Schaeffer 1998) -- for the microlensing events observed towards the Large 
Magellanic Cloud and the Galactic bulge.  Some authors have gone so far as to postulate the existence
of a ``Galactic shroud" (a 2 kpc scaleheight population) 
of white dwarfs around the Galaxy (Gates \& Gyuk 2001) -- but it is likely that the need for this population
would be reduced if we postulate a more substantial (i.e., thicker) thin disk than has been 
previously presumed.  For the dark halo to be comprised disproportionately of white dwarfs would require 
a very significant and unexpected upturn in the white dwarf luminosity function at very faint 
magnitudes, with a peak at $M_V=17-18$ (Monet et al.
2000) and possibly fainter (Fuchs \& Jahrei{\ss} 1998).

The history of direct searches for a Population II white dwarf microlensing population includes several
contentious studies.  For example, Ibata et al. (1999) claimed to have detected 2-5 halo white 
dwarfs in the Hubble Deep Field.  This claim, however, 
has since been withdrawn (Richer 2001).  Johnson et al. (2000) argue for the existence of one white 
dwarf in the HDF south and Mendez \& Minniti (2000) argue for up to ten white dwarfs, but proper 
motion or spectroscopic analysis has yet to be applied to these discoveries.  In their astrometric 
analysis of a section of the 
POSS-I and POSS-II plates, Monet et al. (2000) detected only one 
very high proper motion, faint white dwarf of the type expected for the halo, which
places a strong constraint for only a small local population of halo white dwarfs.
Monet et al. also show that it is very unlikely that a large number of halo white dwarfs
were missed in the seminal Luyten Half-Second (LHS) catalogue (Luyten 1979)\footnote{Although, as noted
by the referee and the editor, many white dwarfs in the LHS may not yet have been positively identified 
due to incomplete spectroscopic observations of this catalogue.},
which strengthens Flynn et al.'s (2001) argument that the Ibata et al. HDF white dwarf
density is inconsistent with the LHS.  

The most recent stirring of this debate is the claim by O01 to have detected 38 
cool, halo white dwarfs.  Reid et al. (2001) and Reyle et al. (2001) have argued that the kinematical 
properties of these dwarfs are consistent with stars of the Galactic thick disk according to the most 
recent models.  On the other hand, Hansen (2001) and Koopmans \& Blandford (2001) argue for, 
respectively, thin disk and halo membership of the O01 white dwarfs.  However, both of the latter 
studies assume a thick disk contribution similar to those favored by earlier Galactic structure
studies (e.g., Gilmore \& Reid 1983, Reid \& Majewski 1993) in which the thick disk
contributes 2\% of the local stars.  More recent studies favor a larger local thick disk contribution 
(coupled with smaller scale heights) in the range of 5-10\% (Robin et al. 1996; Siegel et al. 2001; Chen et al. 2001).  
A large local thick disk density could account for the majority of the O01 white dwarfs since they
are an entirely local sample.

Our contribution to the debate over the O01 claim is that our study is complete enough
to show that within a few thin disk scaleheights, the dominant white dwarf population has 
$U$ and $V$ velocity distributions similar to thin disk M dwarfs (Reid et al. 1995).  This does not 
support the hypothesis of Hansen (2001) that the O01 white dwarfs could be the result of an inflated 
velocity dispersion among the thin disk white dwarfs; our results indicate no inflation of the 
Pop I white dwarf velocity dispersion beyond the 
dispersion seen in other Pop I stars with a large fraction of old members\footnote{Although 
the high thin disk scale height we find would imply an inflated $W$ velocity dispersion, this inflation 
is slight and would not be significant
enough to account for the velocity dispersion of the O01 white dwarfs.}.  Hansen's argument was 
motivated by the fact that the age distribution of the O01
white dwarfs is too young for thick disk membership.  The {\it a priori} assumption in this 
argument is that the thick disk should be uniformly old and distinct from the thin disk.  The fact that 
our more astrometrically complete sample shows a normal velocity dispersion of thin disk-like white 
dwarfs implies that the origin of the O01 sample is more likely to be a thick disk that is {\it not} 
uniformly old, as opposed to an inflated thin disk.  If that is the case, O01 would be a significant 
discovery against models that favor any early formation of the thick disk.  More astrometrically 
complete large-area follow-up surveys would allow one to see the transition from an O01-like velocity 
dispersion sample to an old thin disk-like sample as we have here.  They would show if the transition 
is smooth (as expected for top-down disk formation models) or discontinuous (as a result of sudden 
heating of the thin disk).

Unfortunately, our own survey can not test the halo white dwarf density very well in a direct way.
We do not find any candidate halo white dwarfs.  Although stars 5003 and 10347 have
relatively high velocities, their magnitude is more like that expected of the thick disk than halo, and 
in any case these two stars are not the ultracool white dwarfs that have traditionally been sought for 
microlensing candidates.  On the other hand, we probe a very small 
volume.  For white dwarfs at $M_B=15.5$ (the average absolute magnitude of stars in the O01 survey)
we probe a volume of 500 cubic parsecs, which should net only 0.1 halo white dwarfs based on density
normalizations from starcounts surveys (Reid et al. 2001).  If, as the microlensing results suggest, 
white dwarfs account for 20\% of the dark halo, then we would expect two halo white dwarfs in our
sample.  Our failure to detect any (ignoring the possibility of
an extreme radial velocity for one of our dwarf candidates that would result in a 
peculiar velocity totaling that of a halo star)
suggests that any theoretical white dwarf microlenses would have to be fainter than $M_B=15.5$ or 
that the 20\% fractional contribution of white dwarfs to the halo dark mass is too high (or that we
are just ``unlucky").  We note, however, that for very faint dwarfs ($M_B=17$), our 
volume shrinks to a piddling 60 cubic parsecs, so we are not very sensitive to such dim objects.  This 
emphasizes the point that for faint halo 
white dwarf searches, even precise astrometry and deep photometry are no substitute for surveying 
expansive areas of sky.  The wide-angle approach used by Monet et al. (2000) and O01 would seem the 
correct one for detecting {\it halo} white dwarfs (and checking the WD luminosity function), even though the 
results have so far been more prosaic than hoped.

For all Galactic populations, a high proper motion limit such as that in O01
allows investigation of mainly the the high
velocity tail of the white dwarf population.  This permits an enormous amount of freedom in
interpreting the results of these studies because one must define the velocity distribution function
for {\it each} Galactic population based on {\it a priori} assumptions rather than observational 
constraint.  Our study illustrates the usefulness of more
astrometrically complete data sets that can define the complete velocity distribution of thin disk, 
thick disk and halo white dwarfs.  Such definition will allow clearer context of the nature of the high
velocity white dwarfs normally studied.

While we do not detect any halo MACHO candidates, our results still bear
on the interpretation of the microlensing results.  Alcock et al. (2000)
derived the fractional contribution to the dark matter and average mass of
MACHO's based on a variety of Galactic models to account for the
microlensing of known stellar populations.  Their standard model assumed a
thin disk scaleheight of 300 pc based on previous analyses of
main-sequence stars. However, if the scaleheight of the microlensing
material in the thin disk is {\it not} equivalent to the apparent
scaleheight of brighter main sequence stars -- i.e., if it is closer to 450 pc,
as we have derived here for the white dwarfs -- this potentially increases
the column density of foreground lenses by 50\%, resulting in a similar
line-of-sight depth to the ``maximal disk" model explored in Alcock et al.
(their model F). Their analysis showed that such a line-of-sight depth
resulted in a halo with not 20\%, but {\it 40-60\%} of its mass in MACHO's
and lowered the required lens mass from $\sim 0.5 M_{\sun}$ to $\sim
0.15-0.25 M_{\sun}$.  A MACHO halo with these properties would be less
likely to be comprised of white dwarfs.  If the white dwarf thin disk
scaleheight is 450 pc, then white dwarfs can not be the MACHO population.  
This emphasize Alcock et al.'s point that their analyses of the
microlensing events ``still depend heavily on the model of the Milky Way
and LMC."

\section{Discussion}

Under the assumption that all proposed candidates are {\it bona fide} 
white dwarfs, the results of the present RPMD analysis of the Paper I survey has yielded a 
sky density of likely white dwarfs higher by an order of magnitude over the previous
most complete samples.  Our candidates represent the (expected) low velocity component of the disk 
white dwarf population excluded by
previous proper motion searches.  Several of our candidate white dwarfs are fainter and redder than the 
disk and halo MSTO, so that previous photometric (e.g., UV excess) surveys would not easily have 
found them.  

Our results suggest
substantial astrometric and photometric incompleteness in previous surveys.  Although Wood
\& Oswalt (1998) and Mendez \& Ruiz (2001) have shown that incompleteness does not significantly bias 
derivation of the luminosity function or local density (and we concur that the incompleteness we
describe here does not likely bias derivation of the local density), they do note that deriving star 
formation histories from biased samples is highly susceptible to error when the
proper motion errors are large ($>$ 100 mas yr$^{-1}$).
The tenfold gain in completeness in a large area survey with the photometric
depth and astrometric precision of ours would improve the resolution with which
star formation histories could be delineated from the white dwarf luminosity function.  Such a survey, 
however, would require long time baseline observations at good plate scale.  It is possible that many 
repeat observations over the course of the Sloan Digital Sky Survey could provide this level of 
precision for $V > 20$.  Of course, HST can achieve such precision, but not over a large area, while 
the planned FAME and GAIA astrometric missions will deliver the proper motions, but not the depth.

Our distant white dwarf candidates, while a small sample, provide leverage on the density law well
above the Galactic midplane and 
suggest a higher white dwarf scaleheight than typically assumed, where the 
``old disk scaleheight" of 250-350 pc falls at the very low end of the ``reasonable" range of
scaleheights derived from the 
entire candidate white dwarf sample.  It is noted that the lowest luminosity white dwarf candidates
contribute the highest $V'/V'_{max}$ on average, and when they are excluded the derived 
scaleheight is lowered, though it remains high compared to white dwarf studies at brighter 
magnitudes.  That we should find white dwarfs to have higher scaleheights than non-degenerate 
stars seems, at first blush, consistent with white dwarf cooling 
theory: One might expect the proportion of old stars among white dwarfs to be higher than among
unevolved late-type stars and we 
might expect the highest vertical velocity dispersions for the oldest stars due
to secular dynamical heating processes that progressively increase vertical velocity dispersions
of stars with time.  Yet, the notion of a relatively more heated white dwarf population 
appears to be at odds with the actual kinematics we measure for our sample of candidate
white dwarfs.  One might ask what such a dynamically cold population is doing at such large
$z$.\footnote{Note that postulating overestimates of the distances to our stars does not fix the problem 
because moving the stars closer also make them dynamically colder in the derived transverse velocities.}  
We note that our velocity data do not fall along the Str\"omberg asymmetric drift relation 
(Binney \& Merrifield 1998)
expected for secularly heated disk stars.  A larger sample is needed to check this apparent contradiction.

We have found that the method of maximum likelihood provides lower scale heights that are 
more consistent with (although still generally higher than) those found in the existing literature.  
Maximum likelihood is also far less sensitive to small number fluctuations.  We propose use of this 
more elegant method of analysis to supplement or replace $<V/V'_{max}>$ methods in the future, 
especially when dealing with small samples.

Our statements here must be tempered by two shortcomings of our survey.  First of all, it is clear
that a larger sample of faint white dwarf candidates with low velocities (requiring more
precise proper motions) is needed to better constrain the white dwarf scaleheight, and we hope to 
increase our sample when additional fields with similar plate material are 
analyzed.  In addition, spectroscopic confirmation of the present and any future deep samples
of astrometrically identified white dwarfs would provide much 
stronger confidence in the interpretation of our results.  Radial velocities, if obtainable, will
be critical to verifying the kinematics of this sample, but require 6-10 meter class telescopes
to obtain.  It is hoped that these additional
data will help resolve outstanding questions on the spatial distribution of 
disk white dwarfs.

\acknowledgements

We appreciate useful discussions with Neill Reid, Robert Link and Matthew Bershady.  This work was 
partly undertaken while SRM was supported at the Observatories of the Carnegie Institute of Washington 
by Hubble Fellowship Grant number HF-1036.01-92A awarded to the Space 
Telescope Science Institute which is operated by the Association of Universities for Research in 
Astronomy, Inc. for NASA under contract NAS 5-26555.  It has also been supported by NSF CAREER
Award AST-9702521, a David and Lucile Packard Foundation, and a Cottrell Scholar Award from the
Research Corporation.  We also thank the anonymous referee for useful comments.

\clearpage

{\bf Figure Captions}

\figcaption[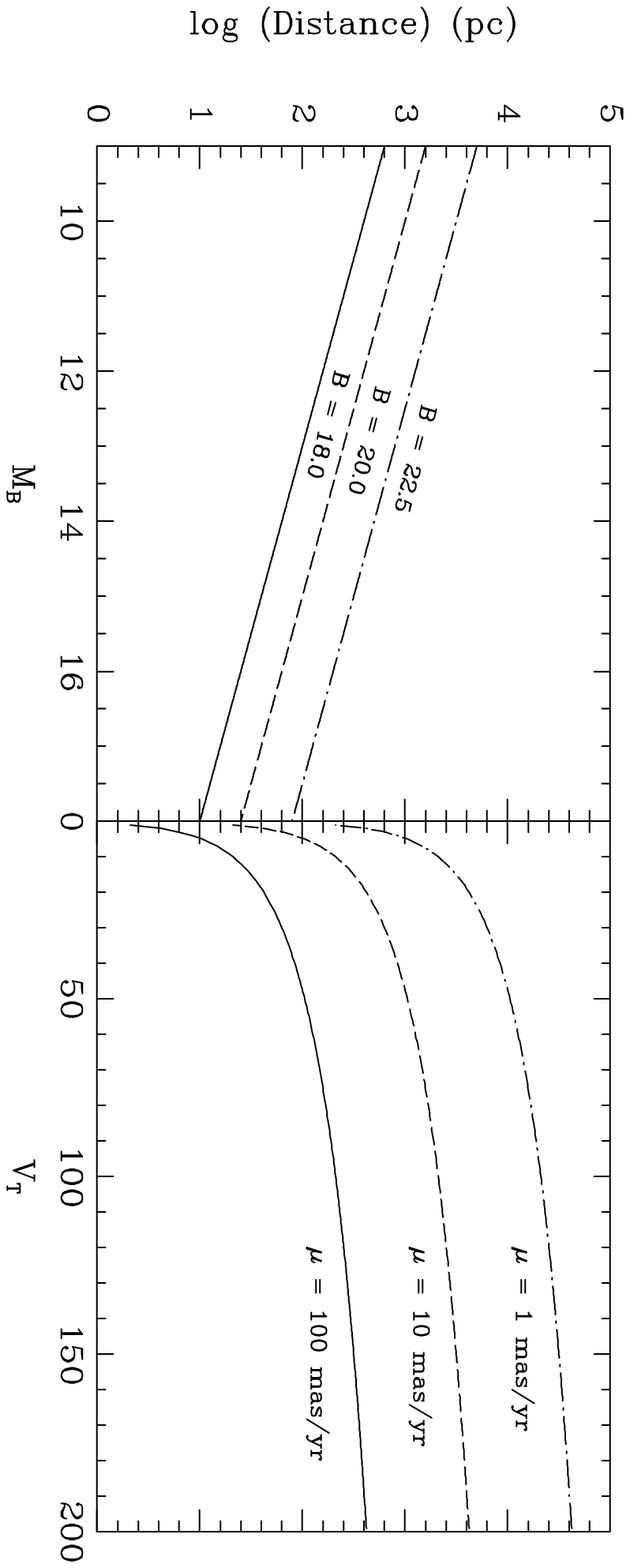]{A demonstration of how photometric (left) and astrometric (right) limits
constrain searchable parameter space.  For any particular limit, observations will
be sensitive to white dwarfs with properties below the corresponding curve.
Note that astrometric limitations restrict the discovery of low transverse 
velocity white dwarfs to relatively short distances.}

\figcaption[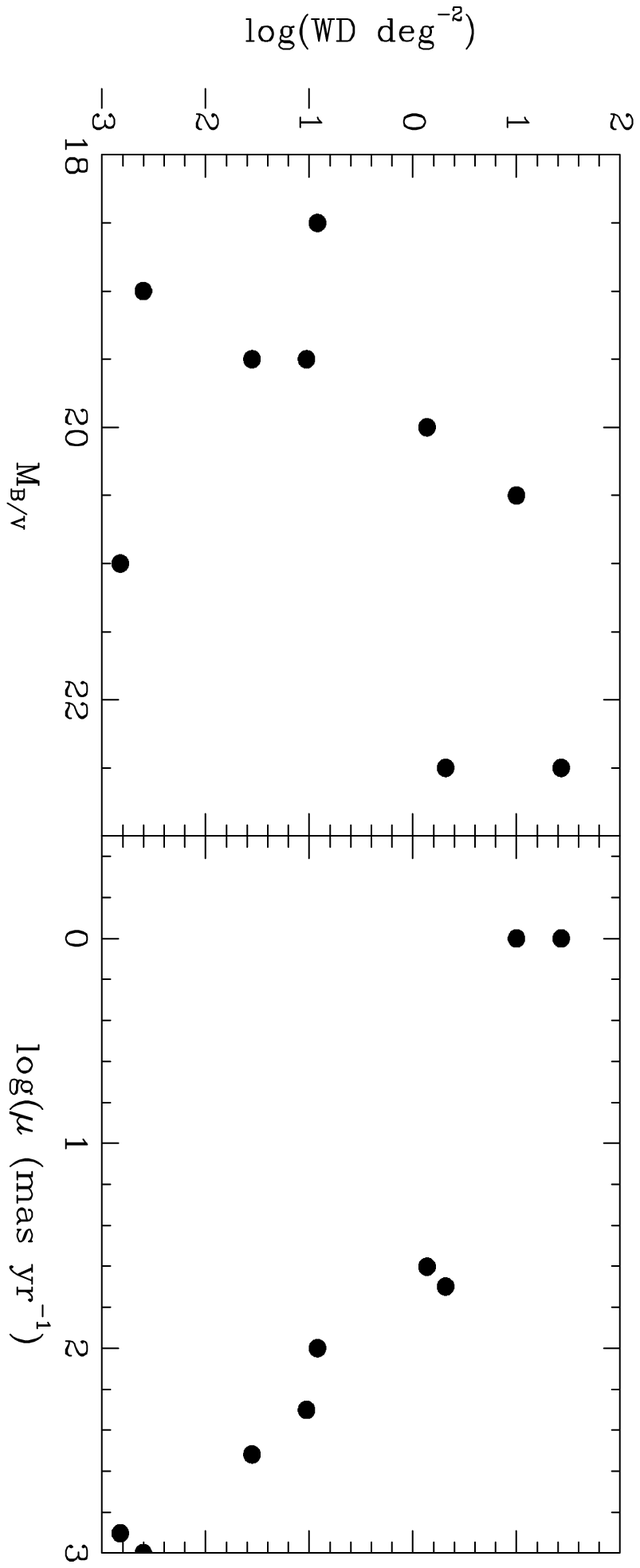]{A demonstration of how the sky density of discovered white dwarfs depends
on photometric (left) and astrometric (right) limitations.  Note the much stronger trend in sky
density with astrometric precision than with photometric depth.  This shows that surveys with greater
astrometric precision will be far more complete than those with superior photometric depth.}

\figcaption[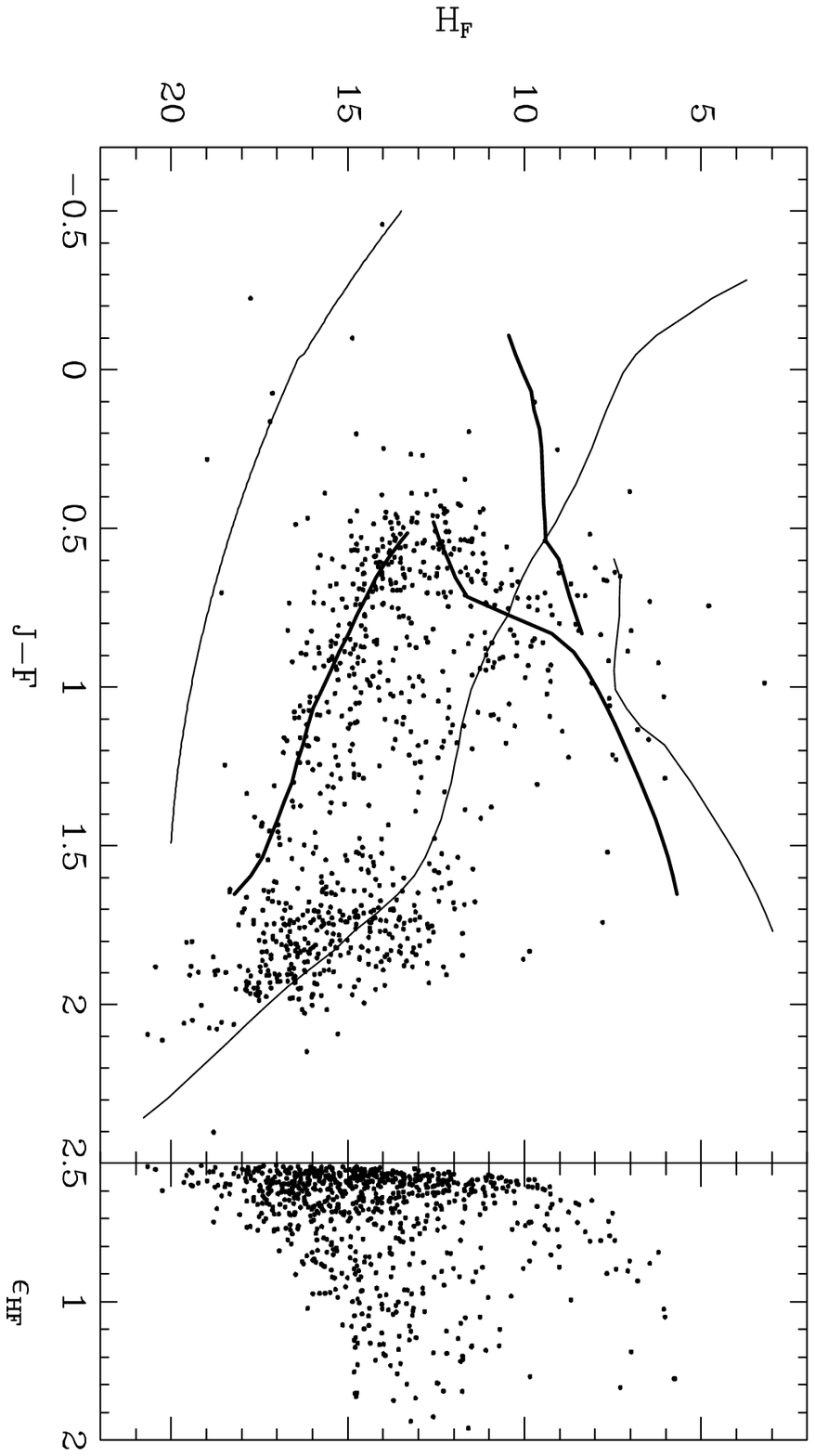]{The reduced proper motion diagram, ($J-F, H_F$) for the Paper I survey.  All spectroscopically 
confirmed QSOs from T94 have been excluded.  Population ridge lines are from Chiu (1980) with the exception
of the WD1 line, which is of our own derivation: Thin lines show the Population I groups MS1, RG1 
and WD1 and thick lines show the Population II groups SD2, RG2 and HB2.  The right panel shows
the distribution of $H_F$ errors and we see that the best $\epsilon_{H_{F}}$ tend to lie at 
high $H_F$, the region of the RPMD of concern in this paper.}

\figcaption[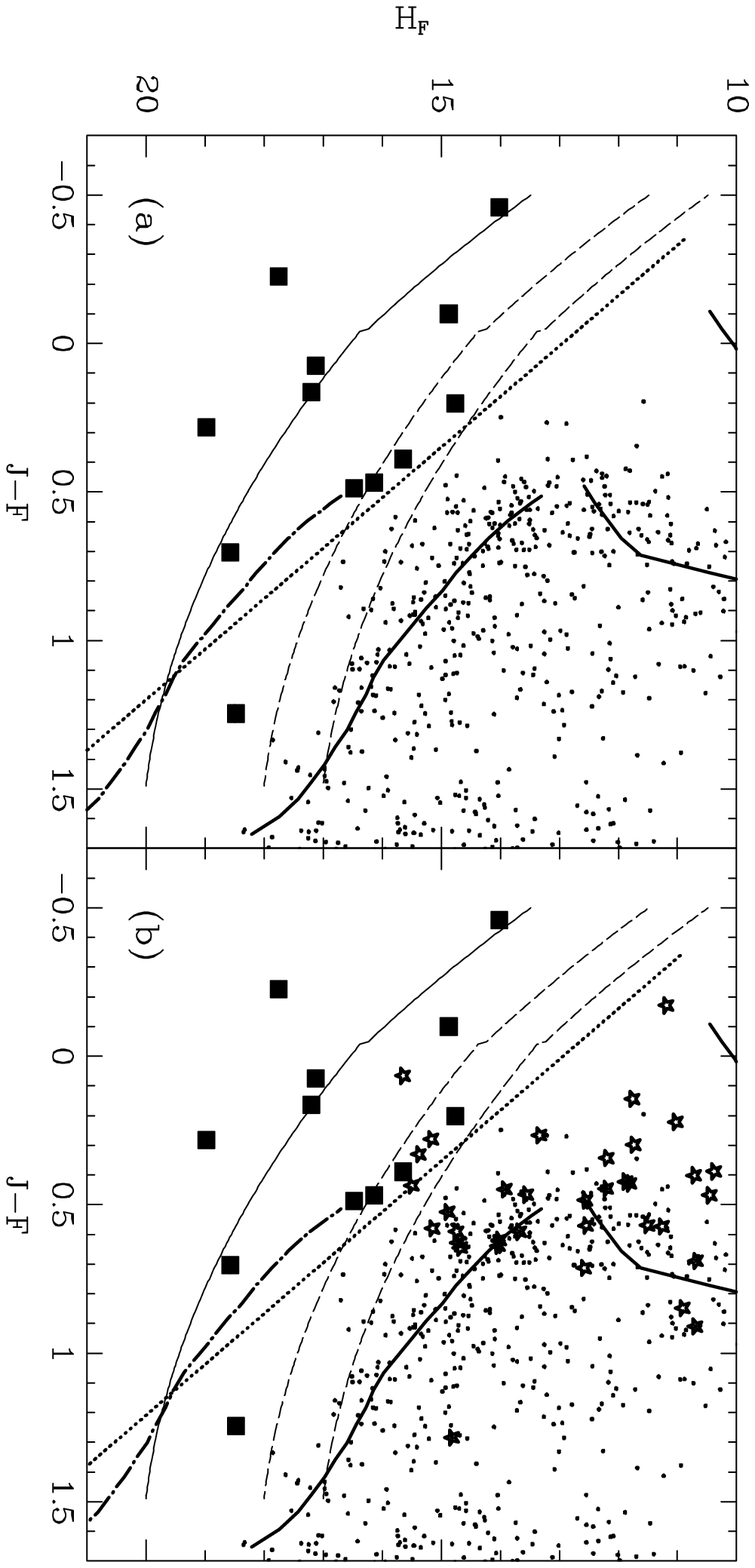]{(a) Enlarged view of Figure 1 highlighting the lower part of the RPMD.  The dotted line shows 
the division between main sequence and white dwarf stars by Jones (1972b).  The dot-dash line
shows the kinematical limit of subdwarf stars.  The dashed lines show the 
75\% and 90\% kinematical bounds for the WD1 as determined by the Population I 5log$V_T$ distribution by Chiu 
(1980).  Solid squares are potential white dwarfs or QSO's.  (b) is identical to (a) but with spectroscopically confirmed QSOs added (starred points).  
Notice how some QSO's inhabit the ``white dwarf" locus of the RPMD.}

\figcaption[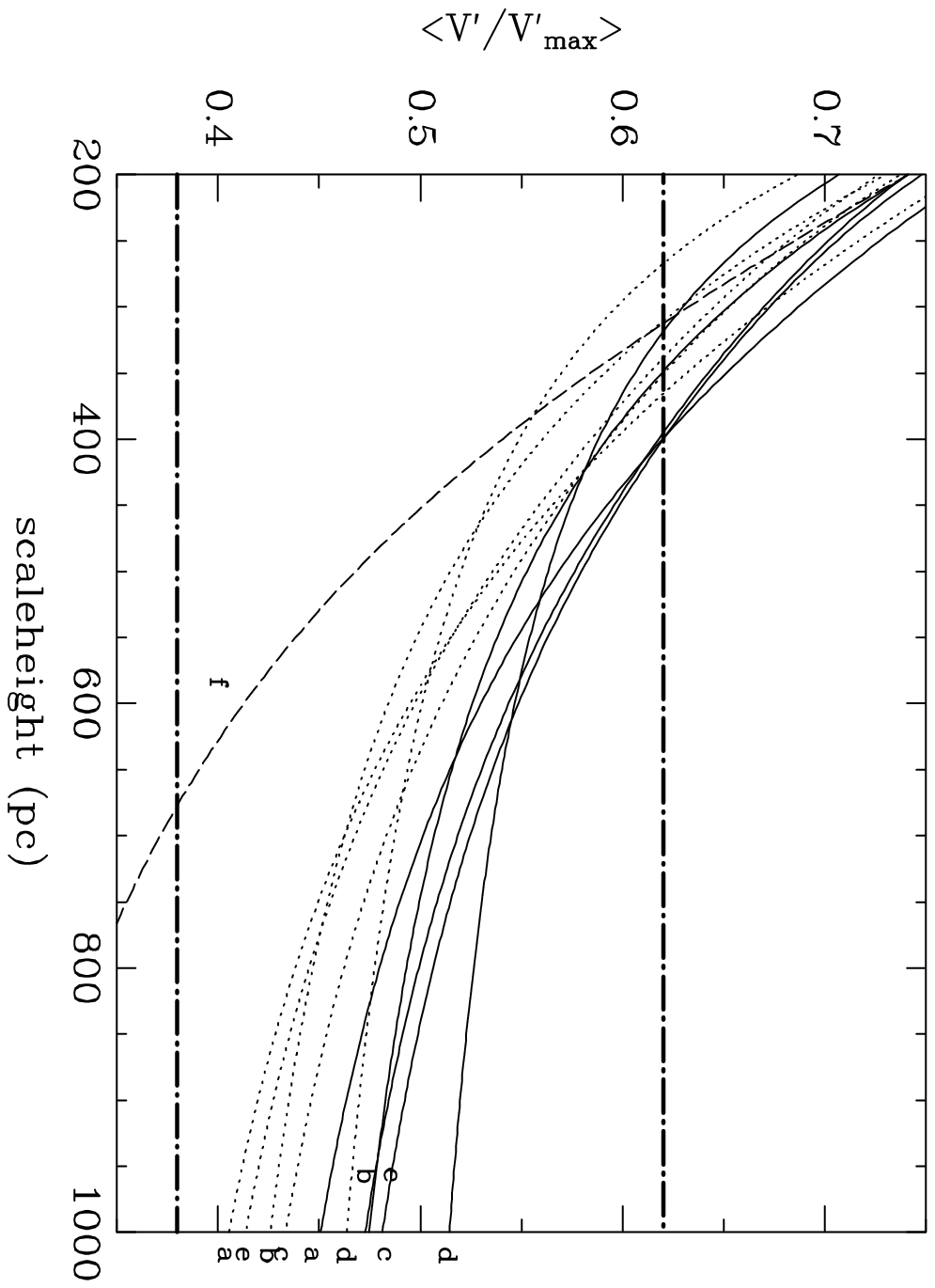]{The mean value of $V'/V'_{max}$  as a function of the adopted scaleheight in the density law.  
The mean value of $V'/V'_{max}$  should be 0.5.  Various subsamples from Table 1 are shown, with curves 
identified in Table 3.  Solid lines are for the subsamples which include the star 17424 while the 
dotted lines are for subsamples which exclude it.  The heavy dot-dash lines indicate the bounds
of $0.38 \leq <V'/V'_{max}> \leq 0.62$, defining the extreme bounds of possible thin disk
scale heights.}

\figcaption[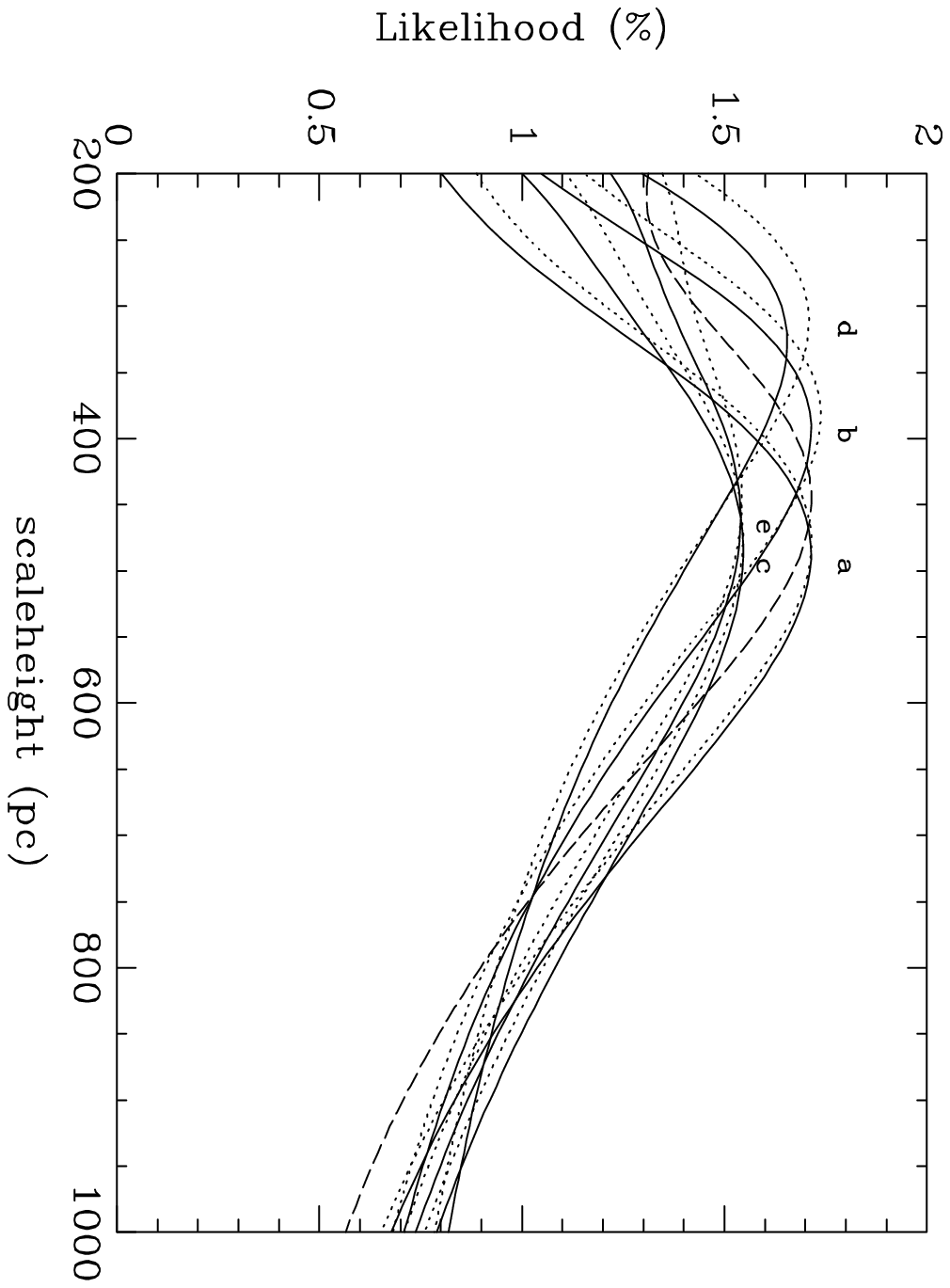]{The relative likelihood for thin disk scale heights.  The likelihood is
integrated in 10 parsec bins.  Notice the tighter clumping of the curves compared to Figure 4  
and the small departure of the dotted lines (those excluded star 17424) from the solid lines 
(those including it), indicating the lesser role that 17424 plays in maximum likelihood
derivations.}
\clearpage

\begin{figure}
\plotone{f1.eps}
\end{figure}
\begin{figure}
\plotone{f2.eps}
\end{figure}
\begin{figure}
\plotone{f3.eps}
\end{figure}
\begin{figure}
\plotone{f4.eps}
\end{figure}
\begin{figure}
\plotone{f5.eps}
\end{figure}
\begin{figure}
\plotone{f6.eps}
\end{figure}

\clearpage

\begin{deluxetable}{lcccc}
\tablecaption{Astrometric White Dwarf Surveys to Date}
\tablehead{
\colhead{Reference} &
\colhead{Area} &
\colhead{Magnitude Limit} &
\colhead{$\mu$ Lower Limit} &
\colhead{Sky Density}\\
\colhead{} &
\colhead{(deg$^2$)} &
\colhead{} &
\colhead{(mas yr$^{-1}$)} &
\colhead{(deg$^{-2}$})}
\startdata
Chiu 1980a              &    0.3 & $V\sim20.0/20.5$ & 1.0\tablenotemark{a} & 10.0 \nl
LDM			&28000   & $B\sim21$   & 800 & .0015\nl
Evans 1992              &   90   & $E\sim20.0$ &  40  & 1.37\nl
Knox et al. 1999        &   28   & $B\sim22.5$ &  50\tablenotemark{b} & 2.07 \nl
Ibata et al. 2000	&  790   & $R\sim19  $ & 1000 & .0025\nl
Scholtz et al. 2000	& 1000   & $B\sim22.5$ & 300 & \tablenotemark{c}\nl
Monet et al. 2000       & 1378   & $B\sim22.5$ & 400 & \tablenotemark{c}\nl
Cooke \& Reid 2000	&   25   & $V\sim18.5$ & 100 & 0.12\nl
Oppenheimer et al. 2001 & 4465   & $R\sim19.5$ & 330 & 0.028\nl
Ruiz \& Bergeron 2001   &  350   & $R\sim19.5$ & 200 & 0.094\nl
This Survey             &    0.3 & $B\sim22.5$ &   1.0 & 27\nl
\enddata
\tablenotetext{a}{See discussion of systematic errors in the Chiu survey in Paper I.  Though
Chiu claims random errors of 1.0 mas yr$^{-1}$, his sample faces several systematic proper
motion errors that reduce the effective proper motion limit of the survey.}
\tablenotetext{b}{Knox et al. claim proper motion accuracies of 10 mas yr$^{-1}$
but restrict their analysis to stars with $\mu > 50$ mas yr$^{-1}$ for $R<20.5$
and $\mu > 60$ mas yr$^{-1}$ for $R\ge20.5$.}
\tablenotetext{c}{These surveys do not state how many of their detection are white dwarfs
and how many are subdwarfs.}
\end{deluxetable}

\begin{deluxetable}{cccccccc}
\tablewidth{0 pt}
\tablecaption{Observed Photometric data for the White Dwarf Candidates}
\tablehead{
\colhead{Name} &
\colhead{$F$} &
\colhead{$\sigma_F$} &
\colhead{$J-F$} &
\colhead{$\sigma_{J-F}$} &
\colhead{$U-J$} &
\colhead{$\sigma_{U-J}$} &
\colhead{v.i.}}
\startdata
\multicolumn{8}{c}{Probable White Dwarfs}\\
\hline
13612	& 17.996 & 0.010 & -0.100 & 0.012 & -0.813  & 0.013   & 3.502\nl
3977	& 20.148 & 0.018 & -0.458 & 0.019 & -1.379  & 0.010   & 1.145\nl
5003	& 20.883 & 0.027 & -0.225 & 0.029 & -1.002  & 0.013   & 1.550\nl
17424	& 21.292 & 0.051 & 1.246  & 0.062 & \nodata & \nodata & 1.291\nl
13786	& 21.317 & 0.066 & 0.163  & 0.067 & -0.621  & 0.033   & 1.073\nl
10517	& 21.620 & 0.153 & 0.703  & 0.154 & -0.454  & 0.309   & 1.174\nl
10347	& 21.654 & 0.041 & 0.282  & 0.050 & -0.689  & 0.165   & 3.916\nl
10405	& 21.971 & 0.090 & 0.074  & 0.095 & -0.949  & 0.035   & 1.576\nl
\hline
\multicolumn{8}{c}{Probable QSOs}\\
\hline
9361    & 21.987 & 0.099 & 0.389  & 0.100 & -0.612  & 0.276   & 0.946\\
4786    & 22.015 & 0.141 & 0.487  & 0.142 & -0.334  & 0.428   & 5.307\\
7701    & 22.139 & 0.070 & 0.202  & 0.075 & -0.750  & 0.232   & 3.131\\
18996   & 22.165 & 0.141 & 0.468  & 0.149 & -0.808  & 0.298   & 2.611\\
\enddata
\end{deluxetable}

\clearpage
\begin{deluxetable}{cccccccccc}
\tablewidth{0 pt}
\tablecaption{Observed Astrometric data for the White Dwarf Candidates}
\tablehead{
\colhead{Name} &
\colhead{$\alpha_{J2000.0}$} &
\colhead{$\delta_{J2000.0}$} &
\colhead{$\mu_{l=0}$\tablenotemark{a}} &
\colhead{$\epsilon(\mu_{l=0})$\tablenotemark{a}} &
\colhead{$\mu_{l=90}$\tablenotemark{a}} &
\colhead{$\epsilon(\mu_{l=90})$\tablenotemark{a}} &
\colhead{$H_F$} &
\colhead{$\epsilon(H_F)$} &
\colhead{p.m.i.}}
\startdata
\multicolumn{10}{c}{Probable White Dwarfs}\\
\hline
13612 &   13:09:56.27 & 29:28:03.8 & -2.78  & 0.95 &  23.57 & 1.03 & 14.87 & 0.13 & 23.07\nl
3977  &   13:09:40.10 & 29:13:35.3 & -1.28  & 0.75 & -5.81  & 0.66 & 14.02 & 0.37 & 8.95\nl
5003  &   13:09:27.40 & 29:15:12.8 & 22.39  & 0.89 & -7.66  & 0.95 & 17.75 & 0.12 & 26.40\nl
17424 &   13:09:31.53 & 29:34:11.0 & -2.87  & 1.93 & -27.21 & 2.02 & 18.48 & 0.23 & 13.55\nl
13786 &   13:09:26.36 & 29:28:21.2 & 13.37  & 1.03 & -6.81  & 0.96 & 17.20 & 0.22 & 14.77\nl
10517 &   13:09:58.45 & 29:23:30.8 & 8.31   & 1.89 & -23.10 & 2.00 & 18.57 & 0.29 & 12.35\nl
10347 &   13:09:13.15 & 29:23:17.1 & 28.37  & 1.61 & -6.93  & 1.56 & 18.98 & 0.17 & 18.17\nl
10405 &   13:08:12.66 & 29:23:23.2 & 8.39   & 1.60 & -6.73  & 1.60 & 17.13 & 0.47 & 6.72\nl
\hline
\multicolumn{10}{c}{Probable QSOs}\\
\hline
9361  &   13:08:35.85 & 29:21:51.2 & 4.43   & 3.02 & -7.05 & 2.75 & 15.65 & 0.10 & 2.97\nl
4786  &   13:09:30.98 & 29:14:50.8 & -3.34  & 2.57 &  0.54 & 2.13 & 16.48 & 0.14 & 2.79\nl
7701  &   13:07:35.05 & 29:19:23.7 & -3.30  & 1.80 & -3.09 & 1.85 & 14.76 & 0.07 & 1.31\nl
18996 &   13:09:09.15 & 29:37:02.2 & 6.10   & 2.95 &  1.22 & 2.57 & 16.13 & 0.14 & 2.12\nl
\enddata
\tablenotetext{a}{Proper Motions in mas yr$^{-1}$}
\end{deluxetable}

\clearpage
\begin{deluxetable}{cccccccccccc}
\tablewidth{0 pt}
\tablecaption{Derived Data for the Probable White Dwarf Candidates}
\tablehead{
\colhead{Name} &
\colhead{$r$} &
\colhead{$\epsilon_r$} &
\colhead{$u$} &
\colhead{$\epsilon_u$} &
\colhead{$v$} &
\colhead{$\epsilon_v$} &
\colhead{$r_{max}$} &
\colhead{$(u^2+v^2)^{1/2}$} &
\colhead{$M_F$} &
\colhead{$M_V$} &
\colhead{$t_{c}$}\\
\colhead{} & \colhead{pc} & \colhead{pc} &
\colhead{km s$^{-1}$} & \colhead{km s$^{-1}$} & \colhead{km s$^{-1}$} &
\colhead{km s$^{-1}$} & \colhead{pc} & \colhead{km s$^{-1}$} &
\colhead{} & \colhead{} & \colhead{Gyr}}
\startdata
13612 & 220  & 50  & -12 & 1  & 36  & 6  & 1700\tablenotemark{a} & 38  & 11.28 & 11.27 & 0.24\nl
3977  & 1630 & 370 & -19 & 6  & -34 & 11 & 4850\tablenotemark{a} & 39  &  9.09 &  8.98 & 0.0014\nl
5003  & 1160 & 270 & 114 & 29 & -31 & 11 & 2660 & 118 & 10.57 & 10.52 & .057\nl
17424 & 180  & 40  & -11 & 2  & -12 & 6  & 198\tablenotemark{b}  & 16  & 15.01 & 15.45 & 6.5 \nl
13786 & 610  & 140 & 30  & 9  & -9  & 5  & 966  & 31  & 12.40 & 12.46 & 0.80 \nl
10517 & 330  & 80  & 4   & 4  & -25 & 9  & 353  & 25  & 14.04 & 14.29 & 2.6 \nl
10347 & 590  & 130 & 70  & 19 & -8  & 6  & 753  & 71  & 12.82 & 12.93 & 1.1 \nl
10405 & 960  & 220 & 29  & 12 & -20 & 10 & 1170 & 35  & 12.06 & 12.10 & 0.62 \nl
\enddata
\tablenotetext{a}{$r_{max}$ limit imposed by astrometric precision.}
\tablenotetext{b}{$r_{max}$ limit imposed by $F$ photometry.  All other limits imposed by $J$
photometry.}
\end{deluxetable}

\clearpage
\begin{deluxetable}{llcccc}
\tablewidth{0 pt}
\tablecaption{White Dwarf Subsamples Used in for $<V'/V'_{max}>$ and $\mathcal{L}$ in Figures 4 and 5}
\tablehead{
\multicolumn{2}{c}{} &
\multicolumn{2}{c}{$<V'/V'_{max}>$=0.5} &
\multicolumn{2}{c}{Max $\mathcal{L}$}\\
\colhead{Curve} &
\colhead{Subsample} &
\colhead{$z_0$\tablenotemark{a}} &
\colhead{$\rho_0$\tablenotemark{b}} &
\colhead{$z_0$\tablenotemark{a}} &
\colhead{$\rho_0$\tablenotemark{b}}}
\startdata
\multicolumn{6}{c}{Solid Curves}\\
\hline
a & all                                               & 705  & 5.3  & 485 & 5.6 \\
b & exclude 3977 (most distant)                       & 745  & 5.2  & 395 & 6.2 \\
c & exclude 5003 (highest velocity)                   & 840  & 5.1  & 485 & 5.6 \\
d & exclude 3977, 5003 (most likely thick disk)       & 1440 & 4.6  & 325 & 6.8 \\
e & exclude 5003, 10347 (two highest velocity)        & 795  & 5.0  & 460 & 5.5 \\
\hline
\multicolumn{6}{c}{Dotted Curves}\\
\hline
a & exclude 17424                                     & 585  & 1.4  & 480 & 1.4\\
b & exclude 17424, 3977 (most distant)                & 545  & 1.4  & 385 & 1.7\\
c & exclude 17424, 5003 (highest velocity)            & 635  & 1.3  & 475 & 1.4\\
d & exclude 17424, 3977, 5003 (most likely thick disk)& 605  & 1.3  & 310 & 2.0\\
e & exclude 17424, 5003, 10347 (two highest velocity) & 595  & 1.2  & 445 & 1.3\\
\hline
\multicolumn{6}{c}{Dashed Curve}\\
\hline
f & 3977, 5003, 10405, 13612, 13786 ($M_V < 12.75$)         &  450 & 0.45 & 445 & 0.36\nl
\enddata
\tablenotetext{a}{Scaleheights are in parsecs.}
\tablenotetext{b}{Densities are in star per 1000 pc$^{-3}$.}
\end{deluxetable}

\end{document}